\documentclass[times,twocolumn,final]{elsarticle}

\usepackage{cnf}

\usepackage{framed,multirow}

\usepackage{amssymb}
\usepackage{latexsym}

\usepackage{url}
\usepackage{xcolor}
\definecolor{newcolor}{rgb}{.8,.349,.1}

\usepackage{hyperref}

\usepackage[switch,pagewise]{lineno} 

\begin{document}

\verso{Yan and Wang}
\journal{}

\begin{frontmatter}

\title{Spatiotemporally Resolved Multi-Scalar Measurements of Methane Tulip Flames in a Square Channel}

\author[1]{Zeyu \snm{Yan}}
\author[1] {Shengkai \snm{Wang} \corref{cor1}}
\cortext[cor1]{Corresponding author}
\emailauthor{sk.wang@pku.edu.cn}{Shengkai Wang}

\address[1]{SKLTCS, CAPT, School of Mechanics and Engineering Science, Peking University, 5 Yiheyuan Road, Beijing, 100871, China}

\begin{abstract}
Understanding the propagation dynamics of premixed flames in confined spaces is important for fire safety in gas pipelines and for optimizing modern internal combustion engines. In sufficiently long channels, premixed flames routinely develop tulip flame structures, yet the dominant mechanism remains elusive, and quantitative data on the evolution of flame morphology and key scalar fields are critically needed to improve the explanation, characterization, and modeling of tulip flame dynamics. In this study, premixed flames of a stoichiometric methane/air mixture were investigated in a square channel at a reduced pressure of approximately 0.3 atm. Time-synchronized, multi-plane, dual-color PLIF measurements yielded a spatiotemporally resolved 3-D dataset of key scalar fields, including temperature and OH concentration, throughout the formation and evolution of the tulip structure. Significant heat loss across the walls counteracted the heat released by combustion, producing a near-constant-pressure environment throughout the experiment. A super-equilibrium distribution of OH concentration was observed in the thermal boundary layers, suggesting that thermal cooling dominated over chemical relaxation in those regions. Additionally, the flame-front morphology at five representative times was determined using a 3-D reconstruction algorithm, from which the flame surface area was extracted. The results of this study should aid theoretical modeling and numerical simulations of premixed flame propagation dynamics in confined spaces under realistic boundary conditions.
\end{abstract}


\end{frontmatter}

\section*{Novelty and significance statement}
This work presents, to the authors’ knowledge, the first quantitative, spatiotemporally resolved measurements of key scalar fields and 3-D morphology of tulip flames traveling in a square channel. Unlike conventional experiments that rely on path-length-integrated measurement methods such as chemiluminescence or Schlieren imaging, this study exploits time-synchronized, multi-plane, dual-color PLIF to resolve flame structures in the lateral direction. The results reveal a significant influence of sidewall heat loss on the scalar distributions, both within the thermal boundary layers and in the bulk flow. The temporal variation of flame surface area was also determined. The new findings and data obtained in the present study promise to advance both theoretical modeling and numerical simulations on premixed flame propagation dynamics in confined spaces and under realistic boundary conditions.

\section{Introduction}
Flames propagating in a laterally confined vessel exhibit dynamics that are distinct from those in free space: confinement by the sidewalls accelerates propagation and causes the flame fronts to elongate along the channel direction and rapidly retract after contacting the sidewalls, leading to a cusp-like inverted flame structure commonly referred to as the “tulip flame” \cite{salamandra1958formation}. This phenomenon is ubiquitously observed in premixed flames traveling in a channel of sufficient length, regardless of the equivalence ratio \cite{dunn1998tulip, yu2015scale, zheng2017experimental, liang2023study}, the type of fuel \cite{starke1986experimental, dunn1998tulip, shen2012experimental, liang2023study}, the specific geometry of the channel \cite{dunn1998tulip, yu2015scale, shen2021phenomenological}(including open-ended \cite{clanet1996tulip, chen2007experimental} or close-ended \cite{starke1986experimental, dunn1998tulip, chen2007experimental, shen2019evolution, shen2021phenomenological} geometry), and the nature of sidewall boundary conditions \cite{gonzalez1992interaction, song2006propagation, li2021numerical, shen2025flame}. The unsteady dynamics of tulip flame propagation are not only physically intriguing, but also practically important to the fire and explosion safety of gas transport pipelines \cite{clanet1996tulip,ciccarelli2008flame, chen2016experimental, shen2019evolution} and to the optimization of advanced internal combustion engines \cite{dorofeev2011flame, xiao2015formation, fan2026three}.

Ever since Ellis’s seminal phenomenological study of tulip flames in the 1920s \cite{ellis1928flame}, there have been a myriad of work on the parametric characterization of their propagation dynamics \cite{starke1986experimental, clanet1996tulip, dunn1998tulip, bychkov2007flame, xiao2012experimental, shen2012experimental, yu2015scale,  zheng2017experimental, shen2019evolution, zhang2019experimental} and on their formation mechanisms \cite{markstein1957shock, guenoche1964flame, starke1986experimental, gonzalez1992interaction, dold1995evolution, clanet1996tulip, marra1996numerical, matalon1997propagation, dunn1998tulip, metzener2001premixed,  xiao2010experimental, ponizy2014tulip, liberman2023dynamics}. Previous experimental measurements of chemiluminescence or Schlieren images and pressure time-histories have yielded a rich collection of data on the global behaviors of flame propagation, such as the flame propagation speed defined by the tip displacement velocity \cite{starke1986experimental, clanet1996tulip, chen2007experimental, ponizy2014tulip, shen2019evolution, liang2023study}, the overall reactivity represented by the total chemiluminescence intensity \cite{yu2015scale, zheng2017experimental}, and the timing of different stages during tulip flame evolution as manifested by the pressure variation \cite{dunn1998tulip, xiao2012experimental, shen2012experimental, zhang2019experimental, xiao2010experimental, shen2021phenomenological}. However, to date, direct experimental studies on the detailed evolution of key scalar fields are relatively scarce.

From a theoretical perspective, a number of mechanisms have been proposed to explain the formation of tulip flames, including: (1) intrinsic flame instabilities of the diffusive-thermal \cite{joulin1979linear, sivashinsky1982instabilities}, Darrieus–Landau \cite{bychkov2007flame, dold1995evolution}, and Rayleigh-Taylor \cite{starke1986experimental, markstein1957shock} types; (2) hydrodynamic mechanisms such as flame-induced flows \cite{ponizy2014tulip, liberman2023dynamics}, burned-gas recirculation \cite{dunn1998tulip, gonzalez1992interaction}, and vorticity generation \cite{matalon1997propagation, metzener2001premixed, xiao2010experimental}; (3) effects of pressure waves \cite{markstein1957shock, guenoche1964flame}; and (4) effects of boundary conditions \cite{clanet1996tulip, ellis1928flame, marra1996numerical}. However, based on the limited experimental evidence available in the literature, the dominant mechanism of tulip flames remains inconclusive.

There are also several studies that aim to reproduce the tulip-flame phenomenon using numerical \cite{gonzalez1992interaction, marra1996numerical, kuzuu1996numerical, song2006propagation, xiao2010experimental, shen2025flame, fan2026three, liberman2023dynamics, zheng2018large} and analytical \cite{dold1995evolution, matalon1997propagation} models. The performance of these models remains to be experimentally validated, particularly with respect to the morphology evolution of flame fronts \cite{bychkov2007flame, shen2025flame, fan2026three} and boundary-layer effects \cite{dunn1998tulip, shen2012experimental, shen2025flame, fan2026three}. To advance the characterization, explanation, and modeling of tulip flame dynamics, quantitative data on flame morphology and key scalar fields, especially in spatially and temporally resolved form, are critically needed.

Historically, spatiotemporally resolved measurements of tulip flame structures and scalar fields were hindered by limitations of the experimental methods used. Conventional measurement techniques, such as chemiluminescence or Schlieren imaging, are inherently path-length-integrated and lack spatial resolution in the lateral direction of flames. Consequently, signals from different cross-sections of the flame overlap, precluding accurate determination of three-dimensional flame morphology and scalar distributions. To address this issue, the current study employs time-synchronized dual-color Planar Laser-Induced Fluorescence (PLIF) of OH radicals \cite{wang2019quantitative, yan2026quantitative} to interrogate the dynamic evolution of tulip flames across selected lateral planes and at different times after ignition. By applying a quantitative spectroscopy model for PLIF signals \cite{yan2025star} and a flame surface reconstruction algorithm, this study also quantifies the 3-D flame morphology, as well as temperature and OH distributions, across different stages of tulip-flame evolution. 

The remainder of this paper is organized as follows: Section 2 describes the current methods of experimental measurement and data analysis; Section 3 presents representative results for methane-air tulip flames under stoichiometric conditions, including flame-propagation dynamics, temperature and OH concentration distributions, and morphology evolution; and Section 4 summarizes the main conclusions and provides an outlook for future studies. 

\section{Experimental Method}

\begin{figure}[ht]
\centering
\includegraphics[width=\linewidth,trim = 0 0 -2 0,clip]{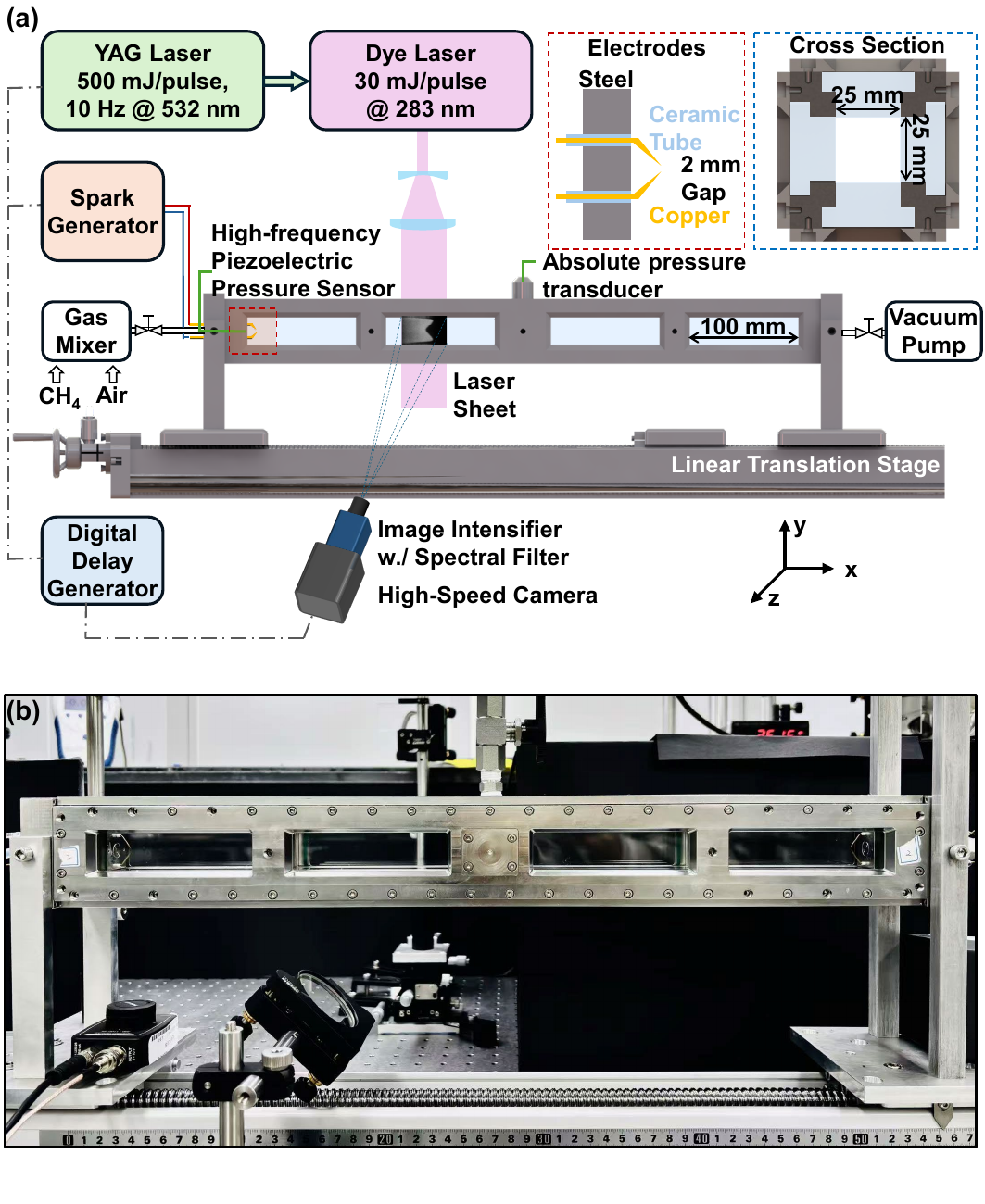}
\caption{Schematic (a) and picture (b) of the current experimental setup.}
\label{fig1}
\end{figure}

\subsection{Optically-Accessible Square Channel and Flow Control System}
As shown in Fig. \ref{fig1}, the current experiments were conducted in a custom-designed closed square channel with an internal cross-section of 25 mm $\times$ 25 mm and a total length of 500 mm. The main frame of the channel was made of SAE 304 stainless steel. To enable optical access throughout the entire flame propagation process, 16 fused silica windows (JGS-1 grade, each 100 mm in length) were flush-mounted along the channel sidewalls. These windows transmitted wavelengths from 185 to 2500 nm, covering the spectral ranges of both the excitation laser and the fluorescence signal.

A mixture of high-purity methane ($99.99\%$) and synthetic air (prepared from $99.999\%$-purity nitrogen and oxygen, with an oxygen mole fraction of 20.85\%) was used. The mixture was prepared manometrically in a 14-L mixing tank to achieve a stoichiometric equivalence ratio ($\phi = 1.0$). During mixture preparation, the gas pressure was monitored by an Omega$^{\rm TM}$ PX01 pressure sensor (with $0.2\%$ accuracy level). To ensure homogeneity, the mixture was mechanically stirred for at least 30 minutes prior to the flame experiments.

A gas inlet and an outlet were mounted on the opposite end walls of the channel and connected to the mixing tank and a vacuum pump, respectively. Before each experiment, the channel was filled with the test mixture to an initial pressure of $P_0$ = 0.32 atm, monitored by an absolute pressure transducer (with $0.2\%$ accuracy level) mounted in the center of the channel. A one-minute settlement period was imposed before flame ignition to minimize flow disturbance. 

Ignition of the premixed gas in the channel was triggered by a single-pulse high-voltage spark generated between two copper electrodes (0.9 mm diameter, 2 mm gap) insulated by alumina ceramic tubes on the inlet wall. The electrodes were powered by a custom-built spark generator with precise timing and energy control. The spark generator comprises an adjustable high-voltage circuit (1–5 kV), a tunable capacitor bank (100–10000 pF), and a TTL-driven low-voltage control circuit. In the current experiments, an ignition energy of approximately 77 mJ was delivered from a 9600-pF capacitor charged to 4 kV, sufficient to ignite the test mixture at low pressures. A Kistler$^{\rm TM}$ 601C high-frequency piezoelectric pressure sensor, also installed on the inlet wall, recorded the pressure variation inside the channel at a sampling rate of 30 kHz.

\subsection{Spatiotemporally Resolved Laser Diagnostics}
High-speed chemiluminescence imaging and dual-color OH planar laser-induced fluorescence (OH-PLIF) were employed to characterize the propagation dynamics, three-dimensional flame morphology, and quantitative scalar-field distribution of the tulip flames. The flame chemiluminescence signal was recorded with a Phantom$^{\rm TM}$ v611 high-speed camera equipped with an EyeiTS$^{\rm TM}$ UV image intensifier at a frame rate of 10 kHz and an exposure time of 99 $\mu$s. The images had a pixel resolution of 1008 $\times$ 512, with each pixel corresponding to a physical size of 0.2 mm $\times$ 0.2 mm, as determined by a geometric calibration process described in the authors’ previous work
\cite{wang2025self, yan2026quantitative}. These images were recorded to obtain the overall evolution dynamics of flame morphology and flame tip position; in addition, they also provided a reference for setting appropriate time delays in the PLIF measurements. 

Under the current experimental conditions, the evolution of tulip flames -- from ignition to flame front elongation, tulip formation and reversion, and transition to steady flame propagation -- spanned a length of approximately two optical windows. Limited by the camera's field of view (slightly longer than one window length), the chemiluminescence images of the two optical windows were recorded separately and combined. As noted in Fig. \ref{fig1}(a), a Cartesian coordinate system was used to define the 3-D positions, with its origin located at the center of the sidewall and the x-axis aligned with the flame-propagation direction.

OH-PLIF measurements were conducted using a dual-wavelength excitation scheme. Specifically, the A-X (1,0) P(1.5)+Q(1.5)+R(2.5) transition cluster near a vacuum wavelength of 282.997 nm and the R(9.5) transition of OH at 282.952 nm were excited by a ns-pulsed tunable dye laser (LIOP-TEC, Model LiopStar-N) at a repetition rate of 10 Hz. The dye laser was optically pumped by a frequency-doubled pulsed Nd:YAG laser (InnoLas, Model SpitLight2000-10) to produce UV pulses near 283 nm. The excitation beam was expanded and shaped by a combination of cylindrical fused silica lenses into a thin sheet of approximately 0.12 mm thickness and 30 mm width. Three representative vertical planes of the channel (located at distances of 0, 5, and 10 mm from the central plane) were illuminated. For flame structures longer than the 30 mm laser sheet width, repeated measurements were performed at multiple horizontal positions, from which the entire flame structure was reconstructed. This was achieved by placing the square channel on a linear translation stage while keeping the positions of the laser excitation and fluorescence collection optics fixed. At the measurement location, the effective pulse energy was approximately 10 mJ, corresponding to a laser fluence of approximately 0.8 J/cm$^2$, which remained well within the linear excitation regime; saturation effects were therefore negligible. The fluorescence signal was recorded with the same intensified high-speed camera used for chemiluminescence imaging, except that a bandpass filter (centered at 310 nm with a bandwidth of $\pm$10 nm) was mounted in front of the lens to block scattered excitation light and isolate the OH fluorescence signal. The intensifier gate width was set to 10 ns -- long enough to cover the entire laser pulse width while short enough to suppress the flame emission background. Further details of the present OH-PLIF diagnostic system can be found in our previous work \cite{yan2026quantitative}.

A multi-channel digital delay generator was used to synchronize the entire measurement system with timing accuracy better than 50 microseconds. A reference signal from the laser control panel served as the master trigger for the delay generator. Once triggered, the delay generator simultaneously sent trigger signals to the spark generator, the intensifier gate, and the data acquisition system for the piezoelectric pressure sensor, after an adjustable delay time $\Delta t$. Measurements were made at different times after flame ignition by varying $\Delta t$. Specifically, three representative moments of tulip flame evolution were selected for investigation: (a) onset of flame-front inversion at 13 ms, (b) maximum global stretch at 19 ms, and (c) restoration of a nearly flat flame front at approximately 23 ms. Based on data from three vertical planes (z = 0, 5, and 10 mm), the three-dimensional structure of the flame surface was reconstructed using a numerical algorithm described in the Supplementary Material. Additional PLIF measurements were also conducted for the central cross-section between 11 and 16 ms.

\begin{figure*}[ht]
\centering
\includegraphics[width=\linewidth]{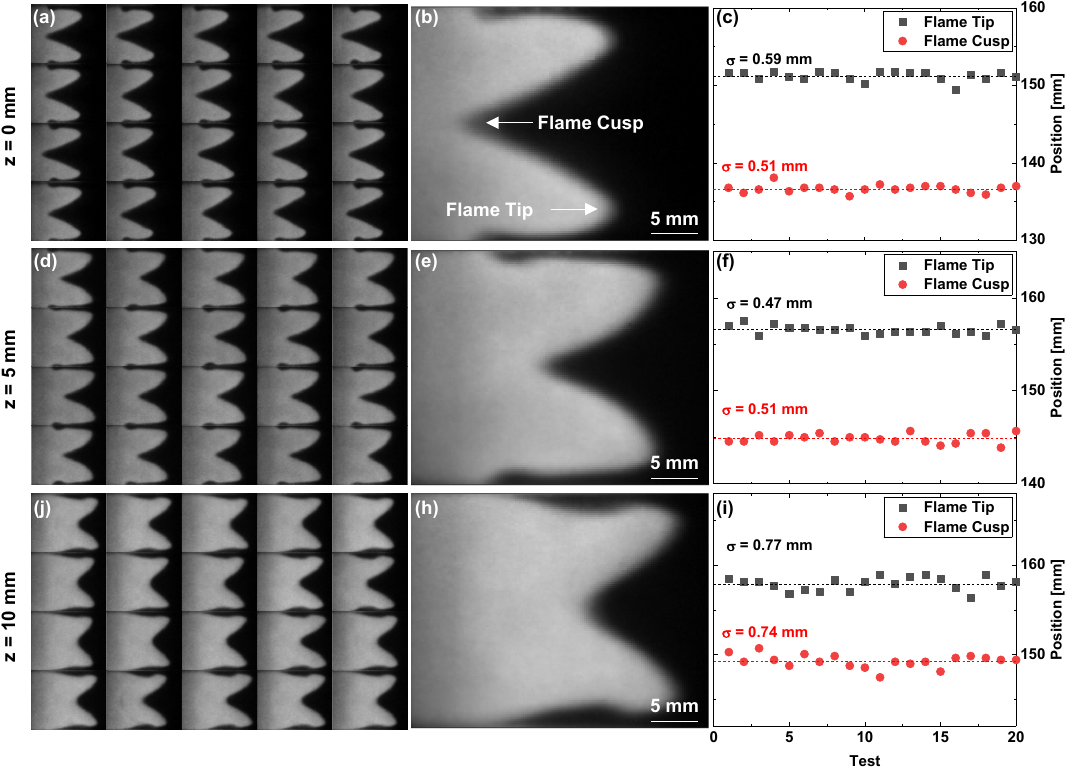}
\caption{Repeatability of transient PLIF measurements at a representative delay time of t = 13 ms. The top, middle, and bottom rows correspond to vertical planes at z = 0, 5, and 10 mm, respectively. The left column displays 20 independent images from single-shot measurements. The middle column shows the corresponding image-averaged results. The right column presents the flame tip and cusp positions extracted from individual OH-PLIF images (symbols) and the averaged results (dashed lines). The standard deviations of the individual flame tip and cusp positions, $\sigma$, are within 0.8 mm (4 pixels).} 
\label{fig_X1}
\end{figure*}

To improve the signal-to-noise ratio and mitigate shot-to-shot jitter of flame ignition, 20 or more PLIF images were acquired for each selected image position, delay time, and wavelength. The resulting PLIF images were averaged within each data group. Given the laminar nature of the tulip flames during their initial evolution stages, the spatial and temporal characteristics of the flames were fairly repeatable under controlled initial conditions\cite{bychkov2007flame, ponizy2014tulip}, as illustrated in Fig. \ref{fig_X1}. For example, the horizontal positions of the flame tip and cusp were found to be relatively stable during repeated measurements, with standard deviations less than 0.8 mm (4 pixels). This averaging suppressed random noise while preserving physical boundaries of the flame front, enabling robust measurements of the 3D flame structures.

\subsection{Quantification of Temperature and OH Concentration Distributions}
The spatial distributions of burned-gas temperature and OH concentration were determined from PLIF images at the two excitation wavelengths, after normalizing the images by the laser intensity distribution obtained from acetone calibration experiments. Further details of the acetone calibration protocol have been documented in the authors’ previous work \cite{yan2026quantitative} and are omitted here for brevity. The temperature was inferred from the intensity ratio ($R$) of the two PLIF signals based on an empirical correlation, $R = 1.48\ \rm{exp}(-2900 K/\textit{T})$, which was also determined in \cite{yan2026quantitative}. The OH concentration was calculated from the PLIF signal at 282.997-nm excitation using a quantitative fluorescence spectroscopy model (StaR-LIF) \cite{yan2025star}. This model properly accounts for the local fluorescence quenching at various temperatures and gas compositions.

\begin{figure}[ht]
\centering
\includegraphics[width=\linewidth,trim = 0 0 -2 0,clip]{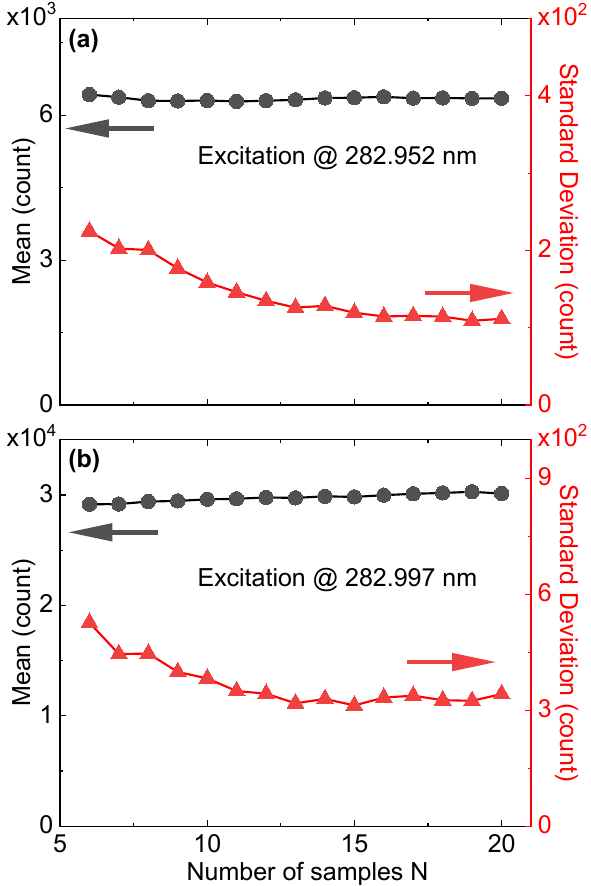}
\caption{Improvement of the signal-to-noise ratio with the number of averages in repeated measurements. Top: 282.952-nm excitation. Bottom: 282.997-nm excitation. Left: mean intensity of the averaged OH-PLIF signal in a relatively uniform region (6 mm $\times$ 6 mm) downstream of the flame front. Right: standard deviation of the averaged OH-PLIF signal in this region.}
\label{fig_X2}
\end{figure}

To evaluate the effect of uncertainty reduction by averaging repeated measurements, the mean and standard deviation of the averaged OH-PLIF intensity in a relatively uniform region (6 mm $\times$ 6 mm) downstream of the flame front are calculated at various number of samples ($N$), as shown in Fig. \ref{fig_X2}. As $N$ increases, the mean intensity remains relatively constant, while the standard deviation of intensity first decays and then converges to a non-zero value. The relative uncertainties in the averaged OH-PLIF signal, defined by the ratio of the standard deviation to the mean intensity, approaches plateau values of approximately 1.8\% and 1.2\% for 282.952-nm and 282.997-nm excitation, respectively.

In the current experiments, the temperature uncertainty, estimated on a root-sum-squared (RSS) basis, was approximately $\pm 3.3\%$ ($\pm 66$ K at a nominal temperature of 2000 K). This uncertainty primarily originated from remaining fluctuations in the averaged PLIF intensity (approximately 1.2\% for the strong wavelength and 1.8 \% for the weak wavelength) and from the uncertainty in the correlation between temperature and the intensity ratio (less than 2.5\%). Accordingly, the total uncertainty for the OH concentration measurement was approximately 15.4\%, arising mainly from intensity fluctuations (1.2\%), uncertainty in the collisional quenching rates in the StaR-LIF model (approximately 15\%), and the propagation of temperature uncertainty (3.3\%).

\section{Results and Discussion}
A 'standard' flame of stoichiometric ($\phi=1.0$) methane/air mixture at sub-atmospheric pressure ($P_0=0.32$ atm) and room temperature was selected as the subject of the current study for the following reasons: (1) the thermal-diffusive instability was minimized at $Le \approx 1$, thus simplifying the flame dynamics and allowing direct interrogation of the hydrodynamic processes commonly observed in tulip flames; (2) the low-pressure environment increased the thicknesses of flame front and quenching layer, enabling fine structures of the flames to be resolved with the current optical diagnostics; and (3) the overall heat release by combustion was roughly balanced by loss across the walls, resulting in a nearly-constant-pressure environment throughout the measurement.

\subsection{Overall Dynamics of Flames Propagating in the Squared Channel}
\begin{figure*}[ht]
\centering
\includegraphics[width=\linewidth]{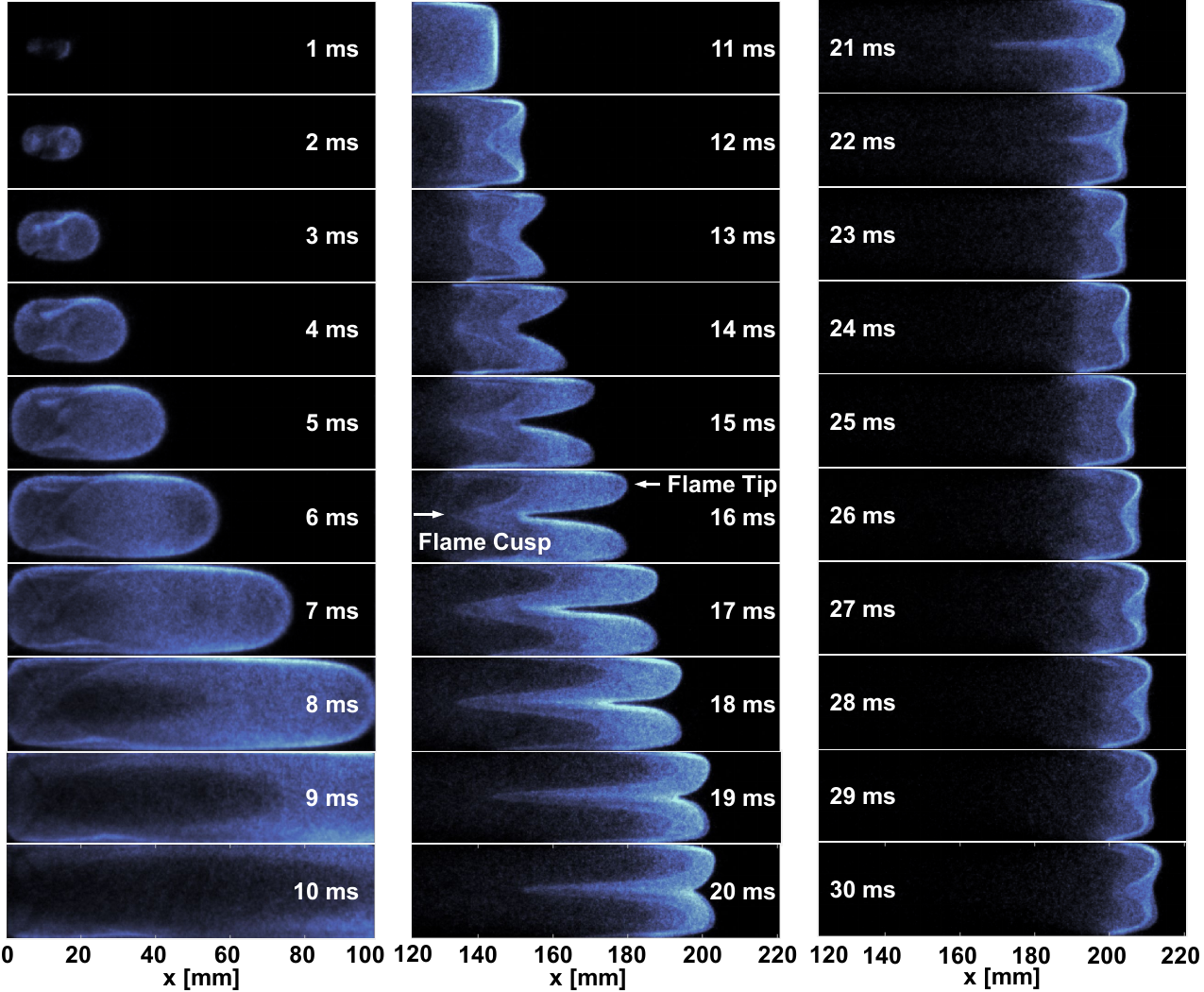}
\caption{Chemiluminescence images of a flame propagating in the square channel showing transitions between finger-shaped, flat, and tulip-shaped flame structures. Experimental conditions: stoichiometric methane/air mixture ($\phi=1.0$), initial pressure $P_0=0.3$ atm. The time interval between frames is 1 ms.} 
\label{fig2}
\end{figure*}

Fig. \ref{fig2} presents a representative image sequence of high-speed chemiluminescence at 1 ms intervals. The flame propagation dynamics exhibit distinct stages as indicated by changes in the flame structure. 

\begin{figure}[ht]
\centering
\includegraphics[width=\linewidth,trim = 0 0 -2 0,clip]{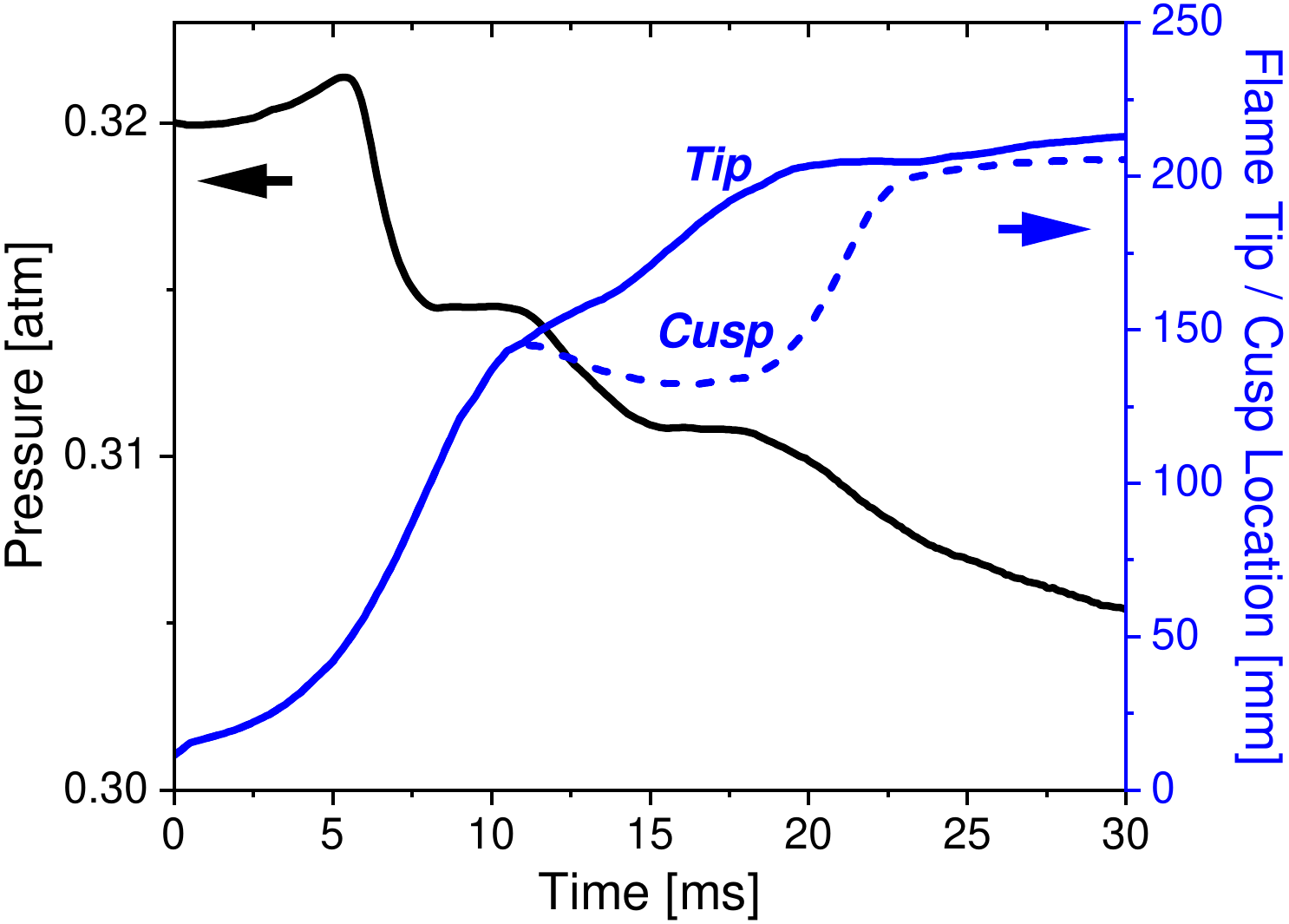}
\caption{Time-histories of the gas pressure (black solid line) and the flame tip (blue solid line) and cusp (blue dashed line) positions.}
\label{fig3}
\end{figure}

Immediately after the onset of ignition (defined as time zero), a flame kernel of approximately 10 mm diameter formed and expanded outward. Due to flow confinement by the endwall and sidewalls, the flame front elongated along the axial direction of the channel, leading to a convex, finger-shaped structure. The rear side of the flame reached the endwall at approximately t = 6 ms and was thermally quenched. The cooling effect of the endwall generated a rarefaction wave traveling forward, quenching the chemiluminescence signal along its path. This endwall contact event was also manifested by a sudden drop in gas pressure, as illustrated in Fig. \ref{fig3}. 

At a time between 10 and 12 ms, the flame skirt also contacted the channel sidewalls, leading to a second pressure drop as shown in Fig. \ref{fig3}. This event was accompanied by flattening of the flame tip region, with a nearly planar flame front occurring at approximately 11 ms and a tip location of approximately 145 mm. These observations agree qualitatively with the previous results of Clanet et al. \cite{clanet1996tulip}, although the exact time and distance of flame propagation differ due to variations in the experimental conditions.

After this point, the flame front receded in the center region while the outer region continued to propagate forward, forming a tulip-shaped structure with curvature inversion around a cusp along the symmetry axis. Recession of the flame cusp was observed up to 16 ms, where it reached a local minimum distance of approximately 133 mm from the channel endwall, and the flame cusp moved forward again after 16 ms, quickly catching up with the flame tip around 23 ms. The greatest overall stretch of the flame, defined as the cusp-to-tip distance reaching its maximum value of approximately 62 mm, occurred at approximately 19 ms. The corresponding change in flame surface area led to a transient increase in the heat release rate that counteracted heat loss across channel sidewalls, yielding a relatively flat pressure trace between 15 and 18 ms. At times beyond 23 ms, the flame front appeared relatively flat (with a minor inversion in the center) as it propagated downstream at a nearly constant speed of approximately 1 m/s; meanwhile, the gas pressure continued to decrease but at a much slower rate. 

\begin{figure}[ht]
\centering
\includegraphics[width=\linewidth,trim = 0 0 -2 0,clip]{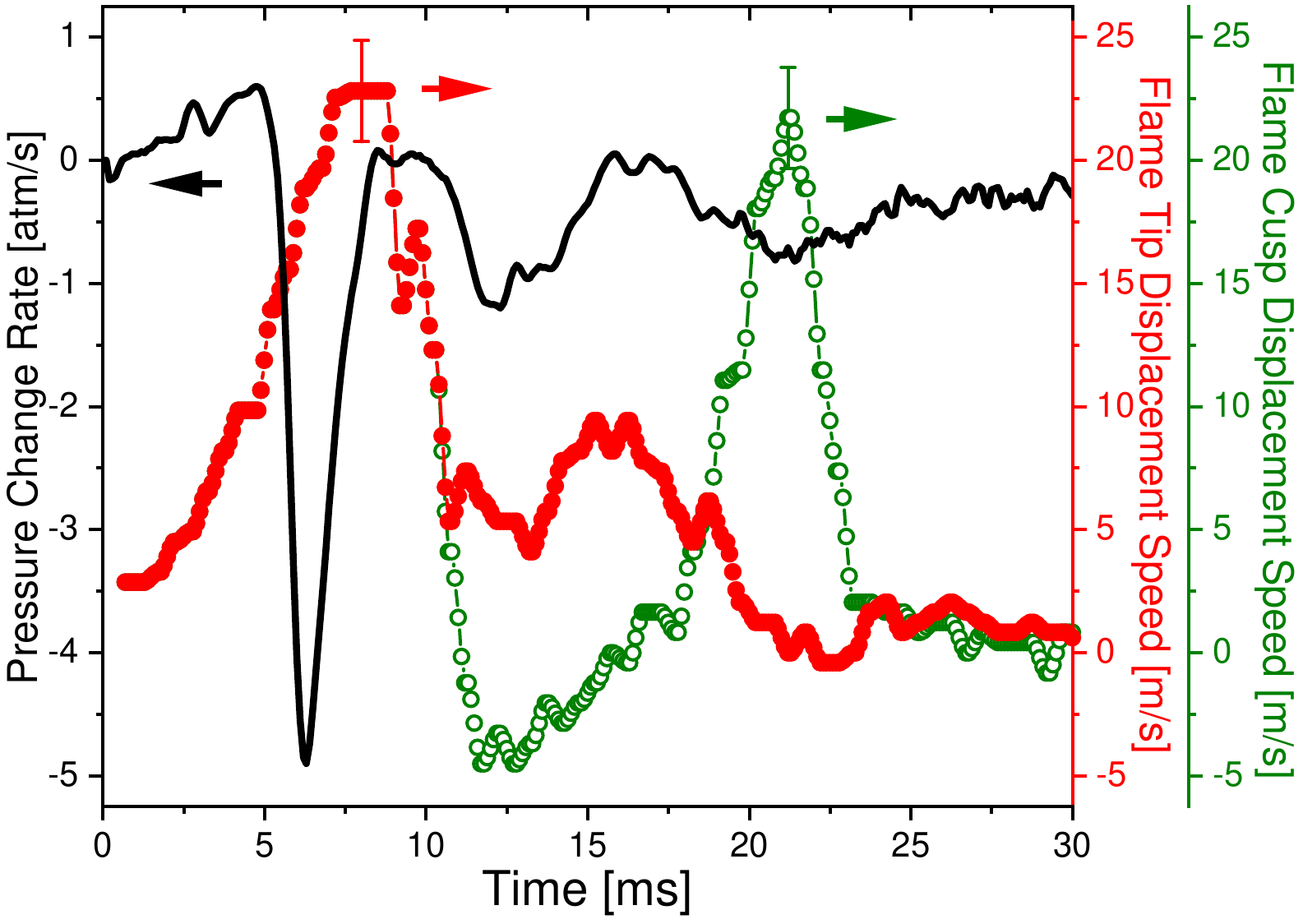}
\caption{Time-histories of the flame tip and cusp displacement speeds, shown together with the pressure change rate. Error bars denote the 2$\sigma$ uncertainty.}
\label{fig4}
\end{figure}

Further details of the overall flame propagation dynamics can be visualized from the displacement velocities of flame tip and cusp, as shown in Fig. \ref{fig4}. The maximum displacement velocities of the flame tip and cusp are very close (23 $\pm$ 2 m/s and 22 $\pm$ 2 m/s, respectively) but occur at different times after ignition (8 ms and 18 ms, respectively). Throughout the tulip flame evolution, the pressure change rate is strongly correlated with the tip displacement velocity, but shows a weaker correlation with the cusp displacement velocity, suggesting that the pressure change is largely governed by sidewall heat loss.

\subsection{Spatiotemporally Resolved Temperature and OH Concentration Fields}

\begin{figure}[h!]
\centering
\includegraphics[width=\linewidth,trim = 0 0 -2 0,clip]{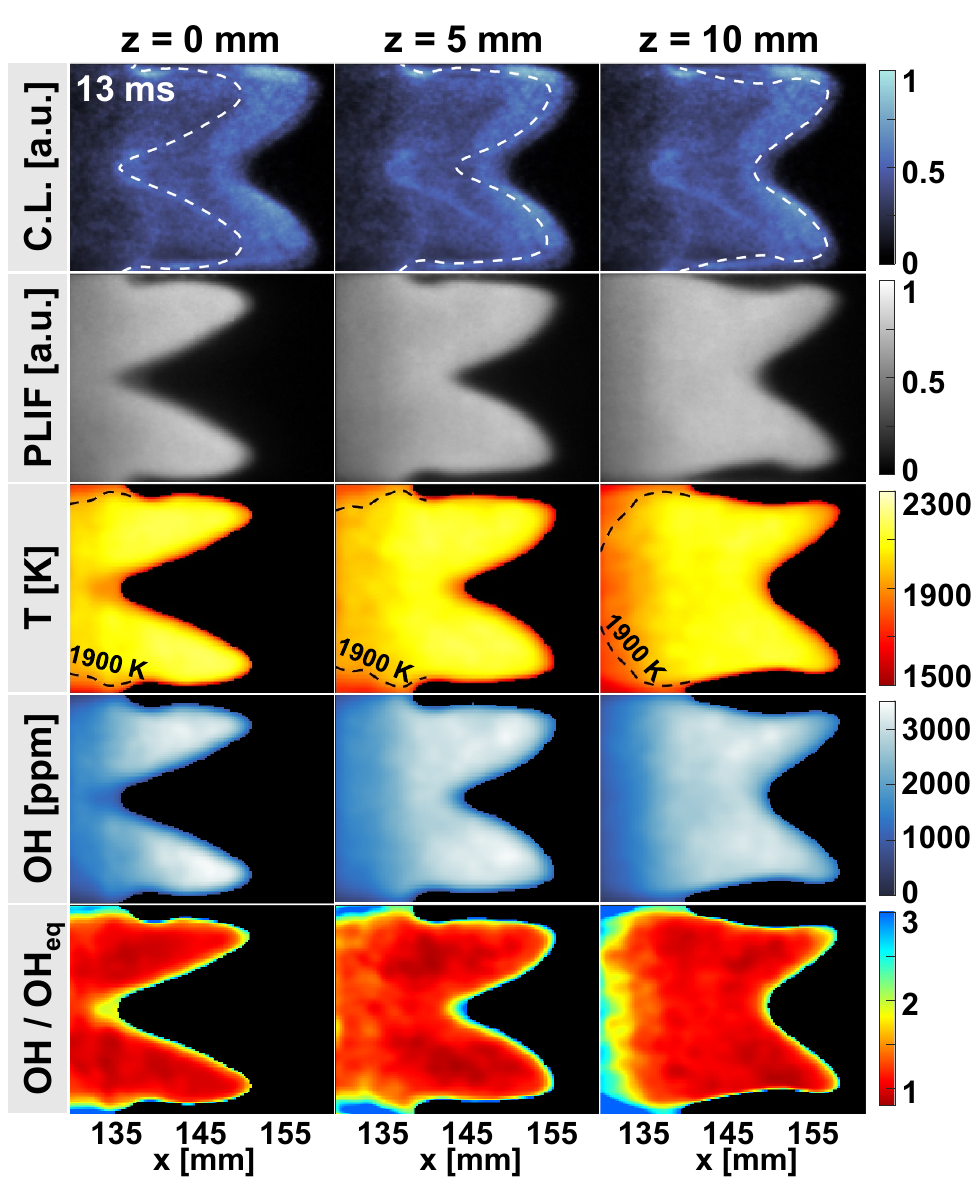}
\caption{Multi-scalar measurements on the z = 0, 5, 10 mm planes at 13 ms. From top to bottom: chemiluminescence (C.L.) signal, PLIF signal at the stronger excitation wavelength, temperature, OH concentration and its ratio to the local equilibrium value. White contours in the top panel indicate the flame front location on the respective plane. A super-equilibrium distribution of OH radicals is evident within the sidewall thermal boundary layers.}
\label{fig5}
\end{figure}

\begin{figure}[h!]
\centering
\includegraphics[width=\linewidth,trim = 0 0 -2 0,clip]{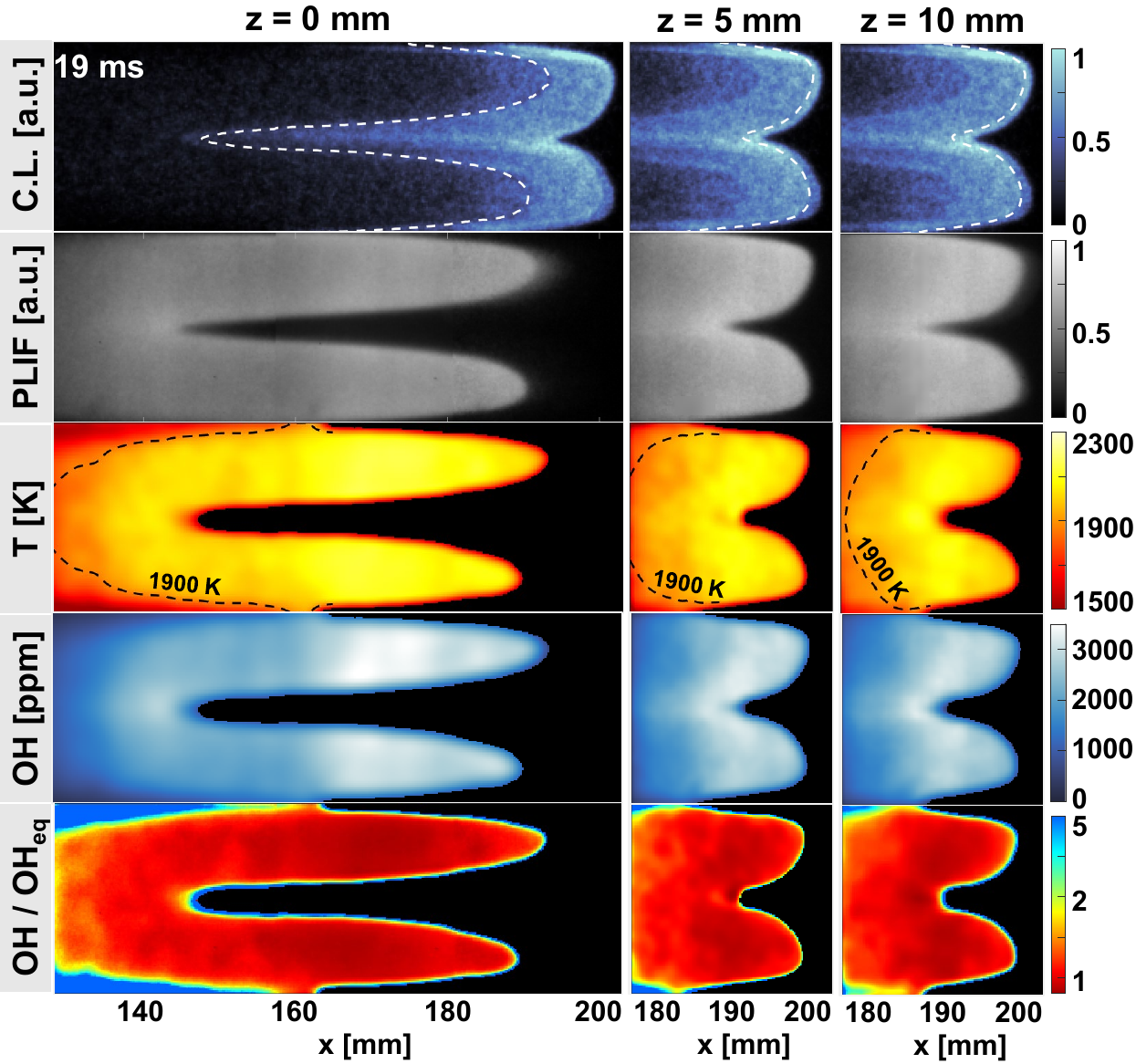}
\caption{Multi-scalar measurements on the z = 0, 5, 10 mm planes at 19 ms, the moment when the overall stretch of the flame front reaches its maximum.}
\label{fig6}
\end{figure}

\begin{figure}[h!]
\centering
\includegraphics[width=\linewidth,trim = 0 0 -2 0,clip]{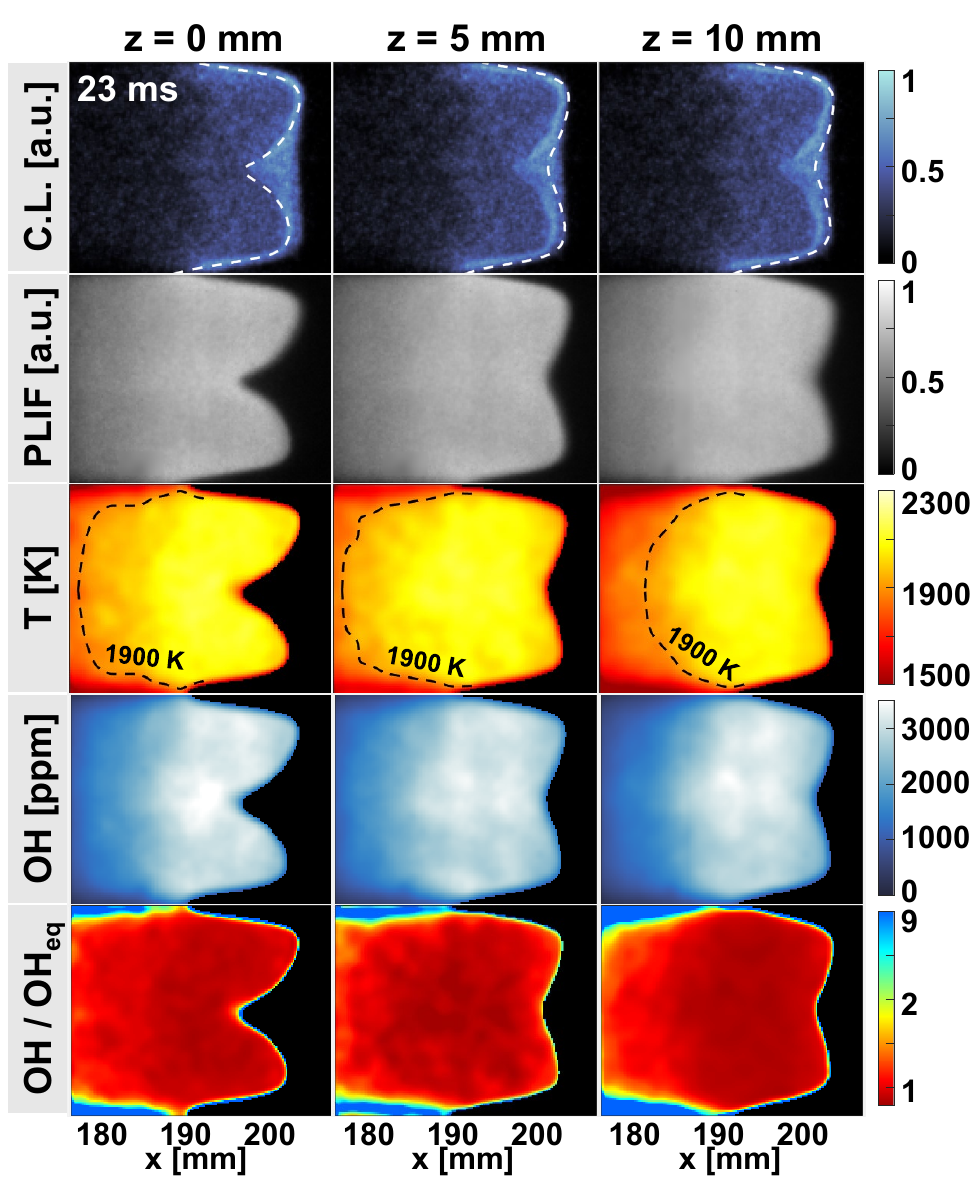}
\caption{Multi-scalar measurements on the z = 0, 5, 10 mm planes at 23 ms, when the flame front recovers a relatively flat shape.}
\label{fig7}
\end{figure}

The spatial distributions of key scalar fields, such as temperature and OH concentration, in the tulip flame at three representative times (13 ms, 19 ms, and 23 ms after ignition) are presented in Figs. \ref{fig5}, \ref{fig6}, and \ref{fig7}, respectively. Results are shown for three vertical planes of the channel, including the central plane (z = 0 mm) and two additional planes located at z = 5 mm and z = 10 mm. Raw images of chemiluminescence and PLIF measurements are also displayed for comparison.

Stratification of the burned gas temperature is observed in all cases -- the temperature immediately behind the flame front is close to the theoretical adiabatic flame temperature ($T_{ad}$  = 2180 K at 0.32 atm), whereas the gas temperature near the sidewalls (denoted as $T_s$) decreases to approximately 1600 K due to heat loss. 

The spatial scale of the near-wall low-temperature region can be quantified by the thermal boundary layer thickness, $\delta_T$, at a streamwise distance $\Delta x$ downstream of the flame-wall contact point. In the current study, this is estimated from an isotherm of $T = (T_{ad} + T_s)/2 \approx$ 1900 K. The choice of other isotherms between 1800 K and 2000 K would yield qualitatively similar results that are scaled by constant factors. For all times ($t$) and lateral positions ($z$) explored in this study, $\delta_T$ is seen to increase with $\Delta x$; moreover, $\delta_T$ also increases with $z$, indicating enhanced thermal cooling near the corners of the channel. The dependence of $\delta_T$ on $t$ appears more complicated and non-monotonic, as a result of unsteady flame propagation.

The near-wall temperature distributions play an important role in the formation of local curvature reversion (concave flame front near the sidewall), as evident in Fig. \ref{fig5}. This phenomenon was not captured by the numerical simulations of Xiao et al. \cite{xiao2010experimental, xiao2014experimental, xiao2015formation} and Shen et al. \cite{shen2021phenomenological}, which assumed adiabatic boundary conditions. The current measurements under realistic thermal boundary conditions suggest that the local curvature reversion is possibly induced by baroclinity due to misalignment between the isotherms and the flame front, which generates vorticity in the clockwise direction near the upper sidewall.

Note that the temperature decrease is not limited to the sidewall boundary layers but extends to the core region behind the flame-wall contact point as well. In this region, the temperature decrease is likely caused by expansion waves originating from sidewall cooling, which have a larger zone of influence than direct heat conduction near the walls.

Also shown in Figs. \ref{fig5}, \ref{fig6}, and \ref{fig7} are the spatiotemporal variations of OH concentration during flame propagation. The measured OH concentration is also compared with the local equilibrium value calculated from the measured temperature and pressure using Cantera \cite{cantera}. They agree reasonably well ahead of the flame-wall contact point; however, within the sidewall thermal boundary layers, the measured OH concentration is generally 3 to 8 times higher than the equilibrium value, suggesting a much faster thermal cooling process than chemical relaxation. This super-equilibrium phenomenon becomes more pronounced at later times as the thermal boundary layers grow thicker.

\begin{figure}[ht]
\centering
\includegraphics[width=\linewidth,trim = 0 0 -2 0,clip]{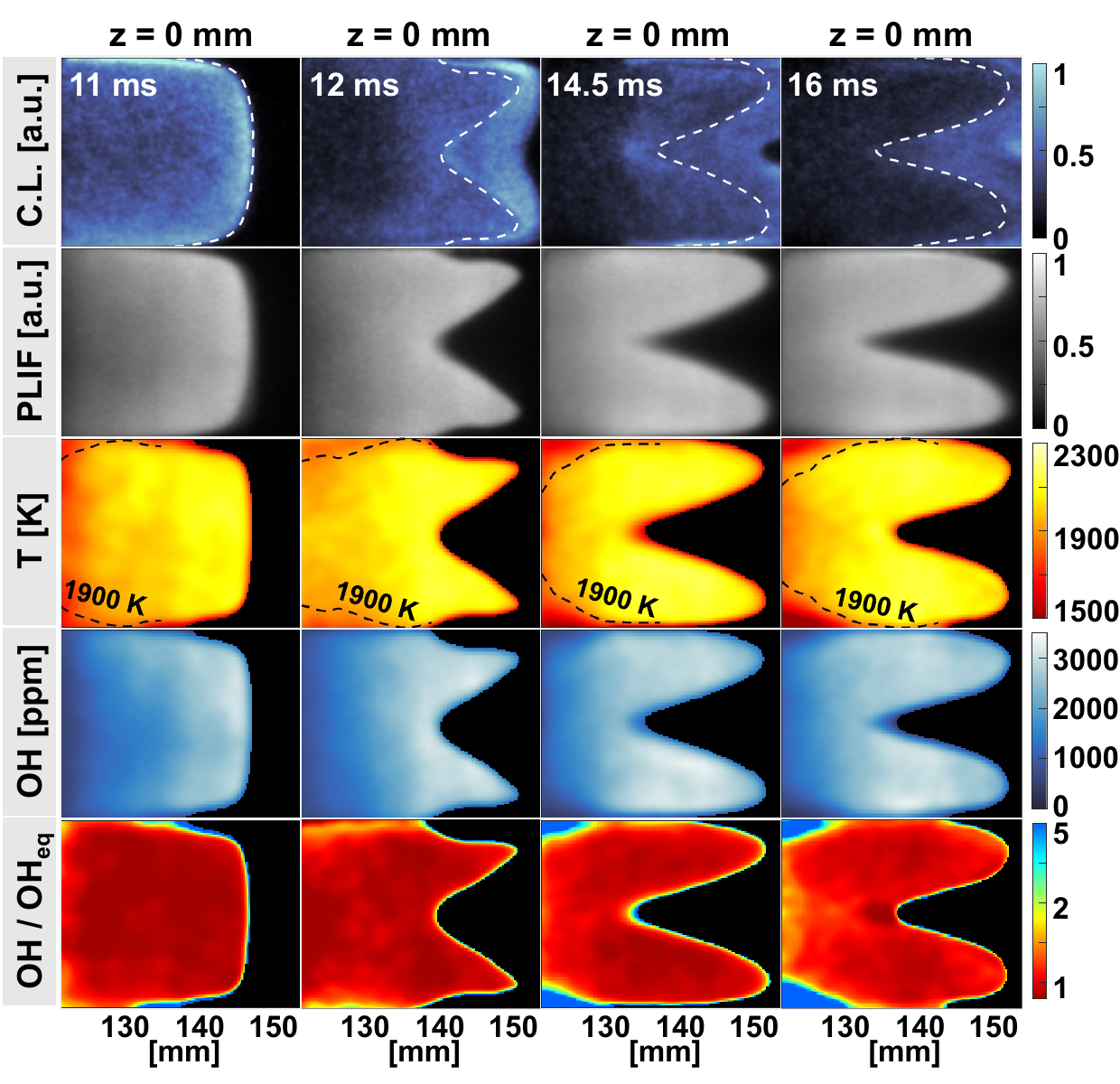}
\caption{Multi-scalar measurements on the central plane (z = 0) at selected times between 11 and 16 ms.}
\label{fig8}
\end{figure}

In addition, the temperature and OH distributions across the center plane at several other times are presented in Fig. \ref{fig8}. The raw data for Figs. \ref{fig5} - \ref{fig8} can be accessed at \cite{Yan2026}.

\subsection{Evolution of 3-D Flame Morphology}

\begin{figure}[ht]
\centering
\includegraphics[width=\linewidth,trim = 0 0 -2 0,clip]{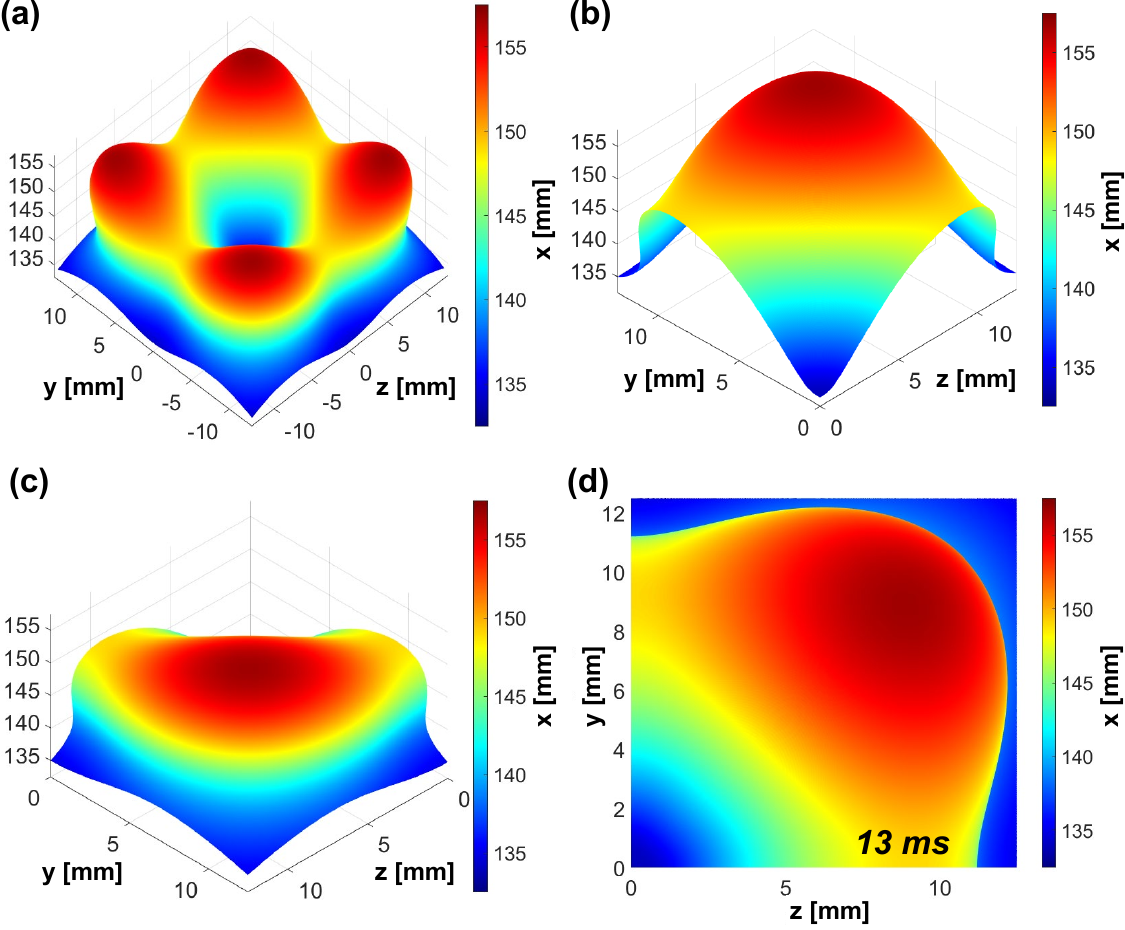}
\caption{3-D structure of the tulip flame front at 13 ms. (a) An overall view of the reconstructed flame front in the square channel. (b)-(d) A quarter flame front viewed from different angles.}
\label{fig9}
\end{figure}

\begin{figure}[ht]
\centering
\includegraphics[width=\linewidth,trim = 0 0 -2 0,clip]{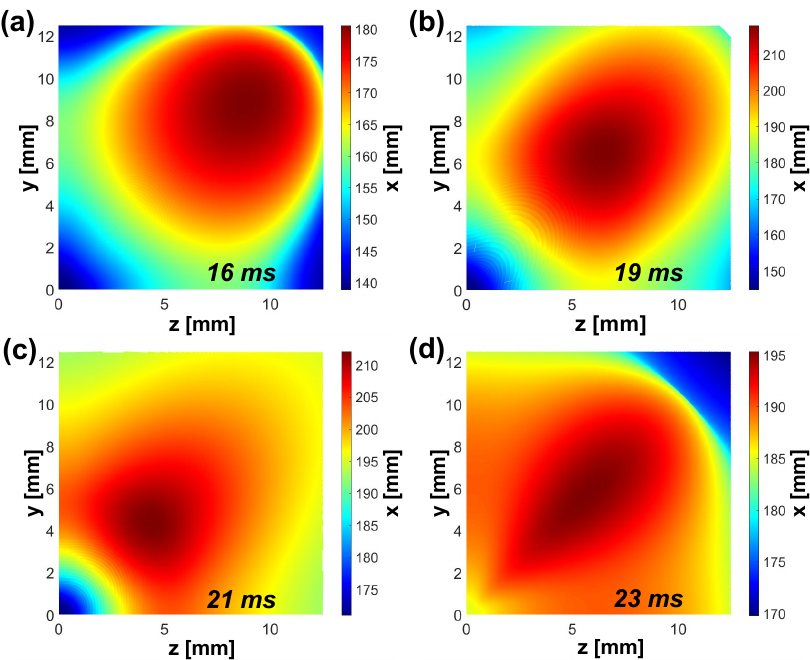}
\caption{Temporal evolution of the tulip flame morphology. (a)-(d) Representative snapshots of a quarter flame front at 16, 19, 21, and 23 ms, respectively.}
\label{fig10}
\end{figure}

For five representative times during the evolution of the tulip flame, the three-dimensional flame front is reconstructed from the PLIF signal of three selected planes (z = 0, 5 and 10 mm) based on symmetry assumptions (see the Supplementary Material for details), as illustrated in Figs.\ref{fig9} and \ref{fig10}. Regarding the overall shape of the inverted flame front and the general trend of its temporal evolution, previous numerical simulations, for example, the LES studies by Shen et al. \cite{shen2025flame} and Zhang et al. \cite{zheng2018large}, agree qualitatively with the present results. From a quantitative perspective, the current results can also provide useful targets for validating and refining numerical models.

Fig. \ref{fig9} presents the reconstructed flame front at $t$ = 13 ms, shortly after the appearance of the tulip structure. Four flame tips are located near the corners of the channel, at a distance of approximately 4 mm from the sidewalls. Between each pair of adjacent flame tips there is a saddle point, located on a symmetry plane and at a distance of 3 - 4 mm from the closest sidewall. Inversion of the flame front (defined by $x \leq x_{avr} \approx 145$ mm) occurs within a lateral distance of 5 mm from the center, and the flame center is recessed from the tips by approximately 20 mm.

As shown in Fig. \ref{fig10}, the evolution of the tulip flame morphology is asymmetric in time. The overall flame stretch (defined by the cusp-to-tip distance) increases from 13 to 19 ms and decreases from 19 to 23 ms, while the region of inverted flame front around the center shrinks monotonically over time. The flame tip locations remain within the region of 5 mm $\le |y|,|z|\le$ 10 mm throughout the elongation and contraction of the tulip flame.

\begin{figure}[ht]
\centering
\includegraphics[width=\linewidth,trim = 0 0 -2 0,clip]{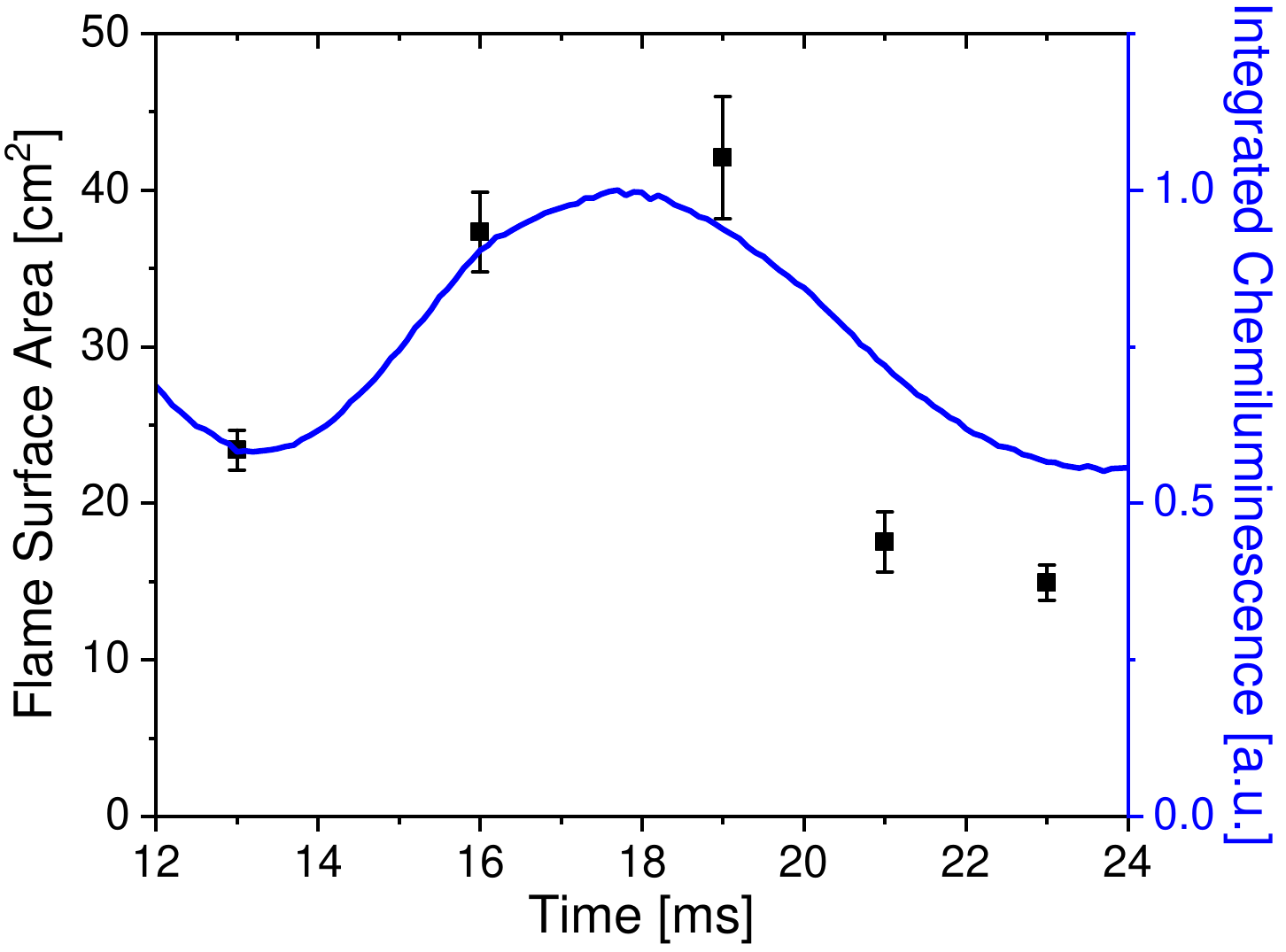}
\caption{Temporal variations of the flame surface area and the integrated chemiluminescence during tulip flame evolution. The integrated chemiluminescence signal was normalized by its maximum value over 12 - 24 ms.}
\label{fig11}
\end{figure}

From the reconstructed flame front, the total surface area of the tulip flame is also determined, as shown in Fig. \ref{fig11}. Throughout the tulip flame evolution from 13 to 23 ms, the ratio of the flame surface area to the channel cross-section area varies between 2.4 and 6.7. Also presented in the figure is the normalized intensity of the spatially integrated chemiluminescence signal, which serves as a proxy for the global heat release rate. The surface area of the flame correlates positively with the chemiluminescence signal, suggesting that the expansion of the flame surface area broadens the effective reaction zone and promotes global heat release. However, the chemiluminescence signal per unit area reaches a minimum around 19 ms when the flame front is mostly elongated, likely due to the sidewall heat loss and quenching over a large flame-wall interaction area. Measurements at longer times ($t$ > 25 ms) are beyond the scope of this study, as gravity-induced asymmetry becomes significant and would complicate the analysis; instead, they are reserved for future investigations.

\section{Conclusions \label{sec: conclusions}}

Spatiotemporally resolved multi-scalar measurements were conducted for tulip flames of a stoichiometric methane/air mixture in a square channel using time-synchronized, multi-plane, dual-color PLIF. A three-dimensional dataset for temperature and OH concentration was obtained at presentative times throughout the evolution of tulip flame structure. The measurements captured the formation of thermal boundary layers near the sidewalls, where the gas temperature dropped below 1600 K. Heat loss across the walls roughly balanced the heat release from combustion, yielding a nearly constant pressure of about 0.3 atm. Within the boundary layers, OH concentration was 3 - 8 times higher than the local equilibrium value, suggesting that thermal cooling occurred much faster than chemical relaxation in these regions. During its propagation and evolution, the tulip flame underwent distinct stages of elongation and contraction. The three-dimensional flame-front morphology, together with the flame surface area, was measured at five selected times throughout this process. These results promise to aid the validation and refinement of numerical models of unsteady flame propagation dynamics under realistic boundary conditions. Notably, although the present study focuses on stoichiometric methane/air flames, the experimental framework is readily applicable to a broad range of fuels and equivalence ratios, which warrants exploration in future studies.

\section*{CrediT authorship contribution statement}
\textbf{Zeyu Yan:} Data Curation, Formal Analysis, Investigation, Writing - Original Draft. \textbf{Shengkai Wang:} Conceptualization, Funding acquisition, Methodology, Supervision, Writing - review and editing.

\section*{Declaration of competing interest}
The authors declare that they have no known competing financial interests or personal relationships that could have appeared to influence the work reported in this paper.

\section*{Acknowledgments}
This work was supported by the National Key Research and Development Program of China under Grant No. 2025YFF0511801, the National Natural Science Foundation of China under Grants No. 12472278 and No. 92152108, and the Space Application System of China Manned Space Program under the project "Ignition Mechanism of Premixed Near-Limit Flames under Microgravity Conditions".

\section*{Supplementary material}
Supplementary Material: Reconstruction of the three-dimensional flame front.

\FloatBarrier

\bibliographystyle{cnf-num}
\bibliography{References}

\end{document}